\documentstyle[psfig]{kapproc} 

\kluwerbib

\startauthorindex

\begin{document}

\articletitle{On the alignment of T\,Tauri stars with the
local magnetic field}

\author{Gaspard Duch\^ene}
\affil{Department of Physics \& Astronomy, UCLA, Los Angeles,
CA 90095-1562, USA}
\email{duchene@astro.ucla.edu}

\author{Francois M\'enard}
\affil{Laboratoire d'Astrophysique, Observatoire de Grenoble,
F-38041 Grenoble cedex 9}
\email{menard@obs.ujf-grenoble.fr}



\begin{abstract}
    Magnetic field is believed to play an important role in
    the collapse of a molecular cloud. In particular, due to
    the properties of magnetic forces, collapse should be
    easier along magnetic field lines, as supported by the
    large-scale sheet-like structure of the Taurus giant
    molecular cloud for instance. Here we investigate whether
    such a prefered orientation for collapse is present at a
    much smaller scale, that of individual objects. We use
    recent high-angular resolution images of T\,Tauri stars
    located in the Taurus star-forming region to find the
    orientation of the symmetry axis of each star+jet+disk
    system and compare it to that of the local magnetic
    field. We find that i) the orientations of the symmetry
    axis of T\,Tauri stars are not random with respect to the
    magnetic field, and ii) that young stars that are
    associated to a jet or an outflow are oriented very
    differently from those which do not have a detected
    outflow. We present some implications of this puzzling new
    result.
\end{abstract}

\section{Introduction}

Stars form in molecular clouds as a consequence of the
gravitational collapse of dense molecular cores. This process
is however sensitive to other phenomena than gravity, such as
rotation or magnetic field. For instance, neutral species are
only affected by gravity while ions are tightly bound to the
magnetic field. Friction between ions and neutrals, known as
ambipolar diffusion, modifies the kinematics of the neutrals,
leading to protostars that are surrounded by pseudo-disks,
even in the absence of rotation (Galli \& Shu, 1993). In the
presence of straight magnetic field lines threading the cloud,
Galli \& Shu (1993) showed that the major axis of the
pseudo-disks are perpendicular to the direction of the
original magnetic field, as expected if the collapse occurs
preferentially along the field lines. The orientation of young
stellar objects (YSOs) can therefore inform us directly on the
importance of magnetic field in the collapse process. In the
event of a leading role for magnetic forces, we expect
circumstellar disks to be oriented so that their
symmetry/rotation axis is (roughly) parallel with the local
magnetic field in the cloud. On the other hand, if magnetic
field does not influence much the collapse, disks might well
be oriented at random.

In the Taurus-Auriga giant molecular cloud, both the dense gas
clouds and the YSOs shows a sheet-like structure fully
consistent with a large scale collapse along the magnetic
field lines (Goodman et al. 1990; Hartmann 2002). At the
individual object scale, Lee \& Myers (1999) showed that
pre-stellar cores were elongated, with an average aspect ratio
of $\sim2.4$. If one attributes this elongation to rotation,
it can be shown that their major axis are also preferentially
found perpendicular to the magnetic field (Hartmann
2002). These findings support the case for a causal link
between the direction of the large scale magnetic field and
the orientation of young stars.

With the advent of high-angular resolution imaging devices, it
is now possible to trace the orientation of already-formed
YZOs. First, collimated jets and outflows from Myr-old
T\,Tauri stars are commonly observed and can be used to trace
their orientation. Strom et al. (1986) noted, during a deep
imaging survey of Taurus sources driving Herbig-Haro objects,
that jets and outflows have a tendency to align parallel with
the local magnetic field. MHD models predict that the ionised
atomic jets of young stars are expelled perpendicular to the
accretion disk (e.g., Shu et al. 1995; Ferreira \& Pelletier
1995), in agreement with striking high resolution observations
of objects like HH\,30 (Burrows et al. 1996). The preferred
jet orientation therefore suggests that the rotation axis of
T\,Tauri circumstellar disks are parallel to the local
magnetic field. A similar conclusion was reached by Tamura \&
Sato (1989) who found linear polarisation vectors of YSOs to
be preferentially parallel to the local magnetic
field. Although the interpretation of this result is affected
by a 90$^o$ ambiguity, it has reinforced the general belief
that the orientation of young stars is in some way related to
the direction of the local magnetic field.

Since early surveys for optical jets from T\,Tauri stars, many
new jets and outflows have been identified. Furthermore, it is
now also possible to spatially resolve circumstellar disks and
accurately estimate their orientation through high angular
resolution imaging. It is therefore time to readress this
issue and study the link between the orientation of the local
magnetic field and the orientation of YSOs in more detail. We
will focus here on the well-studied Taurus star-forming
regions.

\section{Direction of the magnetic field in Taurus}

In the presence of a magnetic field, elongated interstellar
grains tend to spin along a preferred direction, with their
axis of smallest moment of inertia parallel to the magnetic
field (e.g., Lazarian 2002, and references therein). Dichroism
is induced as the grains act like a picket-fence to absorb the
light of background stars more efficiently along the direction
perpendicular to the grains' rotation axis, i.e., generally
perpendicular to the field. A net linear polarisation results
in the other direction, i.e., parallel to the projected
direction of the magnetic field.

\begin{figure}[t]
\hspace*{-0.12\textwidth}\psfig{figure=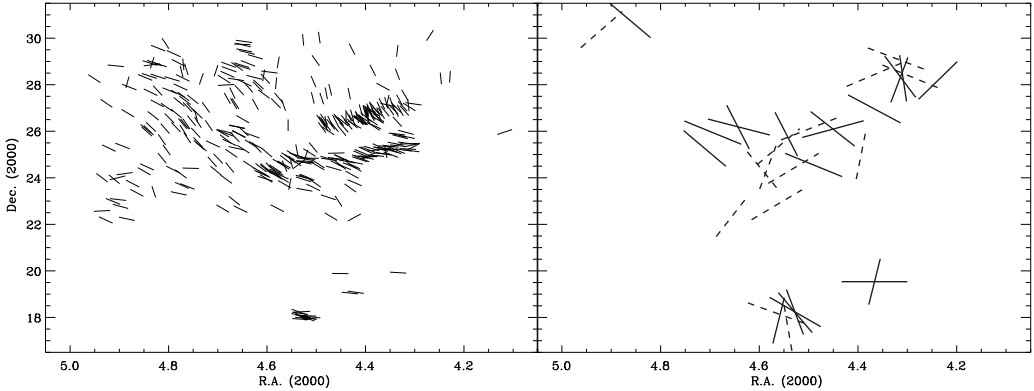,width=1.2\textwidth}
\vspace*{-0.75cm}

\caption{\label{fig:map}{\bf Left:} Polarization measurements
for background stars seen through the Taurus molecular cloud,
tracing the direction of the magnetic field in the cloud. Only
well defined measurements ($P/\sigma_P > 3$) are considered
here. The length of all vectors is uniform and not
proportional to the polarisation rate. {\bf Right:}
Orientation of T\,Tauri stars in Taurus as defined by an
optical jet or outflow (solid vectors) or by a spatially
resolved circumstellar or circumbinary disk (dashed
vectors). In both cases, the symmetry axis of the system is
shown.}
\end{figure}

This technique has been used by numerous authors to infer the
structure of the magnetic field in molecular clouds. For our
analysis, we have compiled over 400 linear polarisation
measurements of background stars in Taurus (left panel in
Fig.\,\ref{fig:map}) from Vrba et al. (1976), Heyer et
al. (1987), Moneti et al. (1987), Tamura et al. (1987), Tamura
\& Sato (1989) and Goodman et al. (1990, 1992). To estimate
the direction of the magnetic field at the location of each
star in our sample (see below), we searched all the
interstellar polarisation data in increasingly larger circles
of radii 0.5--2.0$^o$. We selected the smallest zone
containing at least 4 different measurements of the magnetic
field and retained the median of the measured position angles
as the direction of the local magnetic field. As can be seen
in Fig.\,\ref{fig:map}, the polarisation vectors form a smooth
structure throughout the region and we estimate that the
orientation of the local magnetic field can be estimated to
within 5$^o$ for most objects.

\section{Orientation of T Tauri stars in Taurus}

We have first compiled a complete list of T\,Tauri stars in
the Taurus-Auriga star forming region from the Herbig \& Bell
(1988, hereafter HBC) catalogue. We restrict our study to the
zone 4h00$< \alpha <$5h00 in right ascension and +17$^o <
\delta <$\,+30$^o$ in declination. This yield a sample of over
100 pre-main sequence objects, to which we added HH~30 and
IRAS~04158+2805 (M\'enard et al. 2003), which are considered
as normal T Tauri stars except for their edge-on circumstellar
disk. They are not present in the HBC catalogue because of
their extreme faintness induced by the occulting presence of
their opaque disks. There might be a limited number of
additional objects that we did not include in our study but we
believe that they would not statistically affect our
conclusions.

For each objects, we searched the litterature for the presence
of a spatially resolved jet and/or disk. Only morphological
evidences were used in this process. Collimated jets are
usually identified in deep narrow-band images obtained at the
wavelength of optical forbidden lines (e.g., Mundt et
al. 1991; Dougados et al. 2000) or through long slit
spectroscopy observations, which locate the redshifted and
blueshifted parts of the jet on opposite sides of the star
(e.g., Hirth et al. 1997). In most cases, jets are clearly
resolved and their position angle is known within 10$^o$ or
better. Disks around young stars can be identified in two main
ways: thermal imaging in the submillimeter and millimeter
ranges and scattered light imaging in the optical and
near-infrared. Because current instruments have a limited
dynamic range, the latter technique is usually more sensitive
to edge-on disks (e.g., Burrows et al. 1996) and to
circumbinary rings (e.g., Roddier et al. 1996). In the
millimeter continuum, disks appear as sources of thermal
emission that are spatially resolved when observed with long
baseline interferometers (e.g., Dutrey et al. 1996; Kitamura
et al. 2002). Furthermore, observations of some disks in CO
lines at millimeter wavelength reveal velocity profiles that
are consistent with Keplerian rotation (e.g., Simon et
al. 2000). When available, we used these resolved CO maps to
define the orientation of the disk's semi-major axis. Disk
orientations are known to within 5$^o$ or better when a
Keplerian velocity gradient is detected and within 5--15$^o$
otherwise. More details on the individual sources as well as
data tables are presented elsewhere (M\'enard \& Duch\^ene
2003).

\section{Relative orientations of T\,Tauri stars}

Among our targets, we identified 34 objects with a resolved
jet and/or disk, including 12 that possess both of them. Among
those 12 sources, all but one dubious case (DO\,Tau, see
M\'enard \& Duch\^ene 2003) have perpendicular jets and disks
to within 20$^o$ or better, as illustrated by the prototypical
object HH\,30 (Burrows et al. 1996). The symmetry axis of the
system is determined by the orientation of the jet if it is
present. Otherwise, we assume that the symetry axis lies in a
direction perpendicular to the semi-major axis of the disk. As
for the magnetic field direction, we only find the projection
in the plane of the sky of the symmetry axis of the
object. Despite possible projection effects, which we discuss
below, it is worthwhile searching for possible correlations,
which would trace an underlying physical link between the
orientation of YSOs and the magnetid field.

The orientation of the symmetry axis for the 34 T\,Tauri stars
for which we could determine it is plotted in the right panel
of Fig\,\ref{fig:map} with different symbols for objects with
and without jets. {\it We plot the same axis, i.e. the
symmetry axis of the star+jet+disk system, for both categories
of objects.} Although the direction of the symmetry axes shows
significant scatter, two general trends can be identified from
Fig\,\ref{fig:map}, especially in the central area of the
figure (R.A. $\sim4$h30, Dec. $\sim25^o$). First, T\,Tauri
stars for which a jet or outflow has been spatially resolved
(solid vectors) are majoritarily oriented along the same
direction as the local magnetic field. Second, T\,Tauri stars
without a jet or an outflow but with a spatially resolved
circumstellar disk are roughly oriented perpendicularly to the
local magnetic field. While the former conclusion is similar
to those reached in past studies, the latter point is revealed
here for the first time. We now quantify it in more detail
before exploring its implication for star formation studies.

\begin{figure}[t]
\hspace*{-0.12\textwidth}\psfig{figure=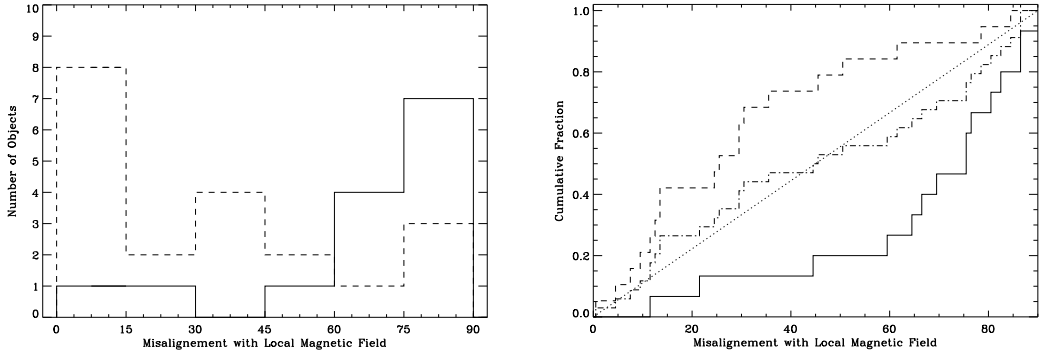,width=1.2\textwidth}
\vspace*{-0.75cm}

\caption{\label{fig:pa}{\bf Left:} Distribution of the angle
between the orientation of the symmetry axis of T\,Tauri stars
and their local magnetic field. From 0$^o$ to 90$^o$, these
two directions go from parallel to perpedicular. The dashed
histogram represent T\,Tauri stars with a jet or an outflow
while the solid histogram are those for which only a disk
could be spatially resolved. {\bf Right:} Cumulative
distributions of angles between the symmetry axis of T\,Tauri
stars and their local magnetic field for objects with a
jet/outflow (dashed), with only a disk (solid) and for the
whole sample of objects with known orientation
(dot-dashed). The dotted line represents the expected
distribution if YSOs were randomly oriented with respect to
the local magnetic field.}
\end{figure}

For each object with a known orientation in the plane of the
sky, we measured the ``misalignment angle'' between the axis
of symmetry of the T\,Tauri star and that of the local
magnetic field. A value of 0$^o$ indicate that both are
parallel while it is 90$^o$ if they are perpendicular. The
distributions of these angles for both our subsamples (objects
with and without a spatially resolved jet) are shown in
Fig.\,\ref{fig:pa} (left panel). The two histograms are
statistically different at the $\sim3\,\sigma$ level: {\it
there is a significant difference in the respective
orientations of T\,Tauri stars with and without jets. Systems
with a jet have their symmetry axis roughly parallel to the
local magnetic field (median angle $=25^o$) while systems
without a jet or an outflow are preferentially perpendicular
to it (median angle $=75^o$).} Interestingly, the distribution
of orientations for our complete sample of 34 objects is
consistent with random orientation, as illustrated in the
right panel of Fig.\,\ref{fig:pa}.

As pointed out earlier, our sample is unlikely to be 100\,\%
complete and an unknown selection effect may be responsible
for this apparent difference. However, the imaging studies
used to build our database of stellar orientations are equally
sensitive to structures (jets, outflows, disks) in any
direction in the plane of the sky. {\it It is difficult to
think of a selection effect that would preferentially sample
stars oriented in a specific direction}. Although there are
strong selection effects regarding inclination, in particular
against pole-on systems, we do not find any significant
difference between the distributions of inclinations of
objects with and without a jet. Therefore, we do not believe
that the distinction between the two subsamples compared here
can be distinguish by specific geometric configurations that
lead to undetectable jets for some YSOs.

\section{Open questions}

The fact that T\,Tauri stars do not orient at random with
respect to the magnetic field and that the presence or absence
of a bipolar jet is statistically linked to the orientation of
the object clearly shows that the magnetic field does play an
important role in the collapse of individual dense cores and
their subsequent evolution. The origin of the differential
orientation trend discovered here is not clear yet and only
suggestions can be put forward at this point. A first
important point to note is that, while jets are most likely
launched through a specific magnetic configuration linking the
star, its accretion disk and the jet, this involves the {\it
stellar magnetic field}. On the other hand, our studies of
orientations concerns the {\it molecular cloud magnetic
field}, which may have a different topology and orientation.

First, it is unclear whether T\,Tauri stars retain the same
orientation throughout their evolution. It could be that the
collapse is strongly driven by the magnetic field, with all
systems having their symmetry axis roughly parallel to the
local magnetic field, and that some of them later become
misaligned through an unknown process. It is worth noting that
the orientation of bipolar outflows from more embedded (i.e.,
younger) Class\,I sources in Taurus show the same tendency to
align parallel to the magnetic field as T\,Tauri stars that
possess a jet (see M\'enard \& Duch\^ene 2003). If all YSOs
retain their original orientation, where are the precursors of
those T\,Tauri stars that now only have a disk?  Despite this
suggestive trend for protostars, we fail to identify any
mechanism that could provide enough torque to rotate by 90$^o$
a star+disk system.

A second possibility is that YSOs are formed with random
orientations, as suggested by the distribution of our complete
sample, and that subsequent evolution lead to the formation of
a powerful jet/outflow only in some conditions. For instance,
Ferreira (1997) showed that a quadrupolar magnetic field
configuration in the disk lead to a much weaker disk-wind, if
any, than a dipolar configuration could achieve. We can then
propose that the configuration of the {\it stellar} magnetic
field will be dipolar if its axis of symmetry is more or less
aligned with the local field in the molecular cloud, while
only quadrupolar (or high order configurations) can be created
if the relative orientation is close to perpendicular. This
could be the consequence of a feedback of the cloud magnetic
field on the (currently unknown) growth mechanism of the
T\,Tauri star field. The fact that Class\,I sources do not
apper to be randomly oriented is somewhat problematic in this
picture though.

A third possible explanation to the observed trend is a
projection effect, in which T\,Tauri stars that have a jet are
located in a different part of the cloud than those who do not
have one. The latter could be seen in projection over the
Taurus cloud, while being located well in front, where the
magnetic field orientation could be different. In that case
however, the field at the periphery of the cloud would need to
be twisted by 90$^o$ with respect to the field found inside
the cloud where T\,Tauri stars with jets would presumably be
located. It is unclear why this would be the
case. Furthermore, one would have to explain why the objects
located in the outer parts of the cloud are systematically
devoid of jets and outflows.

Only tentative interpretation of the observational result
presented here can be given for now; none of them is fully
satisfying. Conducting similar studies in other star-forming
regions, such as $\rho$\,Ophiuchus or Orion, and measuring
submillimeter polarisation vectors from much younger
prestellar cores will help understanding whether this trend is
a general one and possibly identify other factors that play a
role in this preferred alignment of objects with the magnetic
field.

\begin{acknowledgments}
The authors wish to thanks the organizers of the conference
for the quality of its program and its flawless organisation,
as well as Philippe Andr\'e, Bo Reipurth and Hans Zinnecker
for stimulating discussions.
\end{acknowledgments}

\end{document}